\documentclass[aip, rsi, reprint,graphicx]{revtex4-1} 

\usepackage{color}

\newcommand{\lp}{\left(}
\newcommand{\rp}{\right)}

\newcommand{\lsb}{\left[}
\newcommand{\rsb}{\right]}
\newcommand{\lcb}{\left\{}
\newcommand{\rcb}{\right\}}
\newcommand{\labs}{\left|}
\newcommand{\rabs}{\right|}

\newcommand{\abs}[1]{\labs #1 \rabs}

\newcommand{\KAISTNQe}{Department of Nuclear and Quantum Engineering, KAIST, Daejeon 34141, Korea}
\newcommand{\JET}{EUROfusion Consortium, JET, Culham Science Centre, Abingdon, OX14 3DB, UK}
\newcommand{\IPPGreifswald}{Max-Planck-Institut f\"{u}r Plasmaphysik, Teilinstitut Greifswald, D-17491 Greifswald, Germany}
\newcommand{\CCFE}{Euratom/CCFE Association, Culham Science Centre, Abingdon, OX14 3DB, UK}

\newcommand{\ycghim}{\author{Y.-c.~Ghim}\email[]{ycghim@kaist.ac.kr}\affiliation{\KAISTNQe}}
\newcommand{\shkwak}{\author{Sehyun~Kwak}\email[]{slayer313@kaist.ac.kr}\affiliation{\KAISTNQe}}
\newcommand{\jsvensson}{\author{J.~Svensson}\email[]{jakobsemail@gmail.com}\affiliation{\IPPGreifswald}}
\newcommand{\mbrix}{\author{M.~Brix}\affiliation{\CCFE}}
\newcommand{\jetcontributor}{\author{JET Contributors}\thanks{See the Appendix of F. Romanelli et al., Proceedings of the 25th IAEA Fusion Energy Conference 2014, Saint Petersburg, Russia}}

\usepackage{graphicx, amsmath, amsfonts}
\graphicspath{ {images/} }

\newcommand{\calI}{\mathcal{I}}
\newcommand{\calD}{\mathcal{D}}
\newcommand{\vecy}{\mathbf{y}}

\newcommand{\vecO}{\mathbf{0}}
\newcommand{\matK}{\mathbf{K}}
\newcommand{\refeq}[1]{Eq. (\ref{#1})}
\newcommand{\refsec}[1]{Sec. \ref{#1}}
\newcommand{\refapp}[1]{App. \ref{#1}}
\newcommand{\reffig}[1]{Fig. \ref{#1}}

\draft 

\begin{document}


\title{Bayesian modelling of the emission spectrum of the JET Li-BES system} 

\shkwak
\jsvensson
\mbrix
\ycghim
\jetcontributor
\collaboration{\JET}

\date{\today}

\begin{abstract}
A Bayesian model of the emission spectrum of the JET lithium beam has been developed to infer the intensity of the Li I (2p-2s) line radiation and associated uncertainties. The detected spectrum for each channel of the lithium beam emission spectroscopy (Li-BES) system is here modelled by a single Li line modified by an instrumental function, Bremsstrahlung background, instrumental offset, and interference filter curve. Both the instrumental function and the interference filter curve are modelled with non-parametric Gaussian processes. All free parameters of the model, the intensities of the Li line, Bremsstrahlung background, and instrumental offset, are inferred using Bayesian probability theory with a Gaussian likelihood for photon statistics and electronic background noise. The prior distributions of the free parameters are chosen as Gaussians. Given these assumptions, the intensity of the Li line and corresponding uncertainties are analytically available using a Bayesian linear inversion technique. The proposed approach makes it possible to extract the intensity of Li line without doing a separate background subtraction through modulation of the Li beam.
\end{abstract}


\maketitle 

\section{Introduction}
\label{sec:intro}
In fusion research, lithium beam emission spectroscopy (Li-BES) is widely used to measure edge electron density profiles in various machines such as TEXTOR,\cite{92PPCF_Schweinzer, 93RSI_Wolfrum} ASDEX Upgrade\cite{08PPCF_Fischer} and JET.\cite{93PPCF_Pietrzyk, 10RSI_Brix, 12RSI_Brix} When the neutral lithium beam is injected into the plasma both the beam attenuation and emission processes occur due to collisions between lithium atoms and plasma particles. The JET Li-BES system measures the emission spectrum from the spontaneous emission following those collisions. The intensity of the measured Li line depends primarily on the electron density. The relationship between the intensity of the Li line and the electron density can be expressed analytically by a multi-state model,\cite{91APB_Schorn,92PPCF_Schweinzer} which describes excitation and de-excitation reactions caused by the particle impacts including electrons, protons, and impurity ions, and spontaneous emissions.

The Li-BES system is used to infer edge electron density profiles based on intensity profiles of Li line, hence the intensity profiles must be evaluated as precise as possible. Currently, intensity profiles of Li line from JET Li-BES data are obtained via a fitting procedure\cite{10RSI_Brix} with seven fitting parameters: a multiplication factor for background line radiation as measured by beam modulation (one parameter), a quadratic polynomial for the filtered background (three parameters) and a Gaussian function for lithium line radiation with its width, position and intensity (three parameters). 

In this paper we show that improved intensity profiles of Li line can be obtained by modelling both the instrumental function and the interference filter curve for each channel based on Gaussian processes. Bayesian probability theory is used to infer the intensity of the Li line, Bremsstrahlung background, and instrumental offset with associated uncertainties. The instrumental offset can be differentiated from the plasma Bremsstrahlung level since the former is not influenced by the filter function. Thus, the method allows for the separation of signal and background without performing a separate background radiation measurement through a beam modulation procedure. The usage of Gaussian processes to model instrumental effects can be applied generally, as described in \refsec{sec:inference}, to improve spectral fitting also for other systems. A brief overview of the experimental setup of the JET Li-BES system and a description of the measured emission spectrum are given in \refsec{sec:model}. In \refsec{sec:inference} the Bayesian spectral model is described, including the modelling of the instrumental effects with Gaussian processes. The section also shows results from inference on line radiation, Bremsstrahlung background and instrumental effects using these models. A summary is provided in \refsec{sec:summary}.

\section{Spectral modelling}
\label{sec:model} 
\begin{figure}
\includegraphics[width=\linewidth]{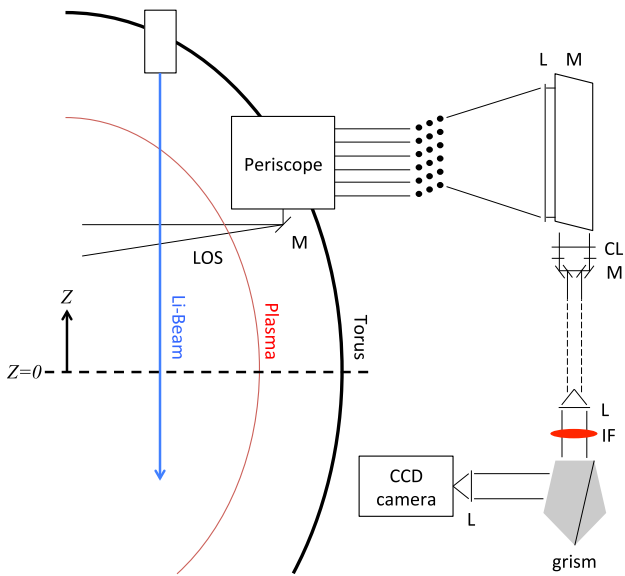}
\caption{Schematic image of the Li-BES system on JET. In the figure L stands for a spherical lens, CL for a cylindrical lens, IF for an interference filter, M for a mirror and grism for a high resolution transmission grating prism. $Z$ is the coordinate of the system where $Z=0$ is at the midplane of JET.}
\label{fig:system}
\end{figure}

The JET Li-BES system consists of 26 spatial channels along the neutral lithium beam, with a typical energy of $\sim55$ keV vertically penetrating into the plasma as shown in \reffig{fig:system}. The 26 line of sights are not perpendicular to the beam direction, causing Doppler shifts of the lithium beam emission. As each line of sight has a different angle to the beam line, the amount of the Doppler shifts are different for different channels. This fact is used to calibrate the spatial positions of the 26 channels. The spectrometer and charge coupled device (CCD) camera are thus required to obtain not only intensities but spectra of the beam emission as well. A detailed description of the system can be found elsewhere.\cite{10RSI_Brix,12RSI_Brix} 

The measured intensity at the $m^\mathrm{th}$ channel denoted as $s_m$ in the CCD camera as a function of $x$ can be expressed as
\begin{equation}
s_m\lp x\rp=d_m\lp x\rp\lsb c_m\lp x\rp a_m+b_m\rsb+z_m,
\label{eq:spectrum}
\end{equation}
where $x$ is the pixel number, corresponding to wavelength, $c_m\lp x\rp$ the instrumental function, $d_m\lp x\rp$ the interference filter curve, $a_m$ the intensity of Li line, $b_m$ Bremsstrahlung background, and $z_m$ instrumental offset. We treat Bremsstrahlung radiation as a constant within a channel because it is almost constant within a narrow wavelength range of $\sim5$ nm set by the interference filter. Note that we measure $s_m\lp x\rp$, and based on this measurement we wish to estimate $a_m$, $b_m$ and $z_m$ among which $a_m$ being the most important quantity, allowing us to infer the local electron density.

Since the spectral width of the Li line is below the resolving capacity of the instrument, the shape of the line on the CCD chip is determined solely by instrumental effects. The filter curve of the system can be independently measured (see in \refsec{sec:inference-system}), and so we separate the instrumental effects into an instrument function $c_m\lp x\rp$, the shape of an infinitely narrow line on the detector, and the interference filter curve $d_m\lp x\rp$.

\section{Bayesian inference with Gaussian processes}
\label{sec:inference}
In order to fit the whole spectrum, the instrumental function $c_m\lp x\rp$ and the interference filter curve $d_m\lp x\rp$, must be known. The functional shape of these are not known a priori, so we here use non-parametric Gaussian processes to model them.

A Gaussian process,\cite{06MIT_Carl} defined on a one dimensional domain, is defined by a covariance function and a mean function, where the covariance function specifies the covariance between any two points in the domain. This restricts the variability of the function to be inferred, and can be used instead of a parameterization. Gaussian processes were introduced in the fusion community in reference\cite{11EFDA_Svensson} and is the default way of representing profile quantities in the Minerva framework.\cite{07ISP_Svensson} It has been used for current tomography,\cite{13NF_Romero} soft-x ray tomography,\cite{13RSI_Li} and for representing profile quantities.\cite{11EPS_Schmuck,15NF_Chilenski} The covariance function of the Gaussian process is usually defined through families of covariance functions where a few so called \textit{hyperparameters}, such as overall scale and length scale of the function, define the shape of the covariance function and thus the variability of the function.

Denoting the instrumental function of the $m^{\mathrm{th}}$ channel $c_m\lp x\rp$, and the interference filter function $d_m\lp x\rp$, as a vector $\vecy_*=\lsb y_{*i} \rsb$ where $y_{*i}$ corresponds to $c_m\lp x_i\rp$ or $d_m\lp x_i\rp$, the representation of these functions as Gaussian processes corresponds to the $y_{*i}$ having a multivariate normal distribution
\begin{equation}
\vecy_*\sim\mathcal{N}(\vec0,\mathbf{K}(\mathbf{x},\mathbf{x})),
\label{eq:GPR}
\end{equation}
so
\begin{eqnarray}
p\lp\vecy_*\rp&=&\frac{1}{\sqrt{\lp2\pi\rp^N\abs{\matK}}}\times
\nonumber\\
&&\exp\lsb-\frac12\lp\vecy_*-\vecO\rp^T\matK^{-1}\lp\mathbf{x},\mathbf{x}\rp\lp\vecy_*-\vecO\rp\rsb.
\label{eq:GPR-explicit}
\end{eqnarray}
Here, $p\lp\vecy_*\rp$ is the probability density function of $\vecy_*$, $N$ is the number of elements in $\vecy_*$ which is the total number of pixels within each channel. $\matK$ is a $N\times N$ covariance matrix whose $ij^\mathrm{th}$ component is determined as
\begin{equation}
K_{ij}=\sigma^2_f\exp{\lp-\frac{1}{2l^2}|x_i-x_j|^2\rp}+\sigma^2_n\delta_{ij},
\label{eq:cov}
\end{equation}
where $\sigma^2_f$ is the signal variance which regulates the overall scale, $\sigma^2_n$ is the noise variance controlling the noise level of the signal, and $l$ is the length scale governing how fast the function can change significantly. $\sigma_f$, $\sigma_n$ and $l$ together are the hyperparameters. $\delta_{ij}$ is the Kronecker delta. To determine the instrumental function, we use Bayesian probability theory with \refeq{eq:GPR-explicit} as the prior, where the hyperparameters are determined based on the measured data by maximizing the evidence as described in \refapp{app:evidence}.

The prior distribution \refeq{eq:GPR-explicit} is then used in Bayes formula
\begin{equation}
p\lp\mathbf{y}_*|\mathbf{y}\rp=\frac{p\lp\mathbf{y}|\mathbf{y}_*\rp p\lp\mathbf{y}_*\rp}{p\lp\mathbf{y}\rp},
\label{eq:Bayes}
\end{equation}
to find the posterior Gaussian process $p\lp\mathbf{y}_*|\mathbf{y}\rp$ where $\vecy$ is the measured data. The likelihood, $p\lp\mathbf{y}|\mathbf{y}_*\rp$, is a probabilistic model of the observations, and includes the noise characteristics. 

In the following section, we provide a detailed description of how the likelihood and the prior are applied for emission spectrum modelling of the JET Li-BES system.

\subsection{Instrumental function and interference filter curve inference}
\label{sec:inference-system}
We infer the instrumental function and filter curve $\vecy_*$, i.e., $c_m\lp x\rp$ or $d_m\lp x\rp$ in \refeq{eq:spectrum}, by maximizing the posterior in \refeq{eq:Bayes} where the prior is defined using Gaussian processes as described above. The instrumental function $c_m\lp x\rp$, can be derived from the emission spectrum data during the beam-into-gas calibration measurements, using separate interference filter curve measurements for $d_m\lp x\rp$. During the beam-into-gas calibration measurement, neutral lithium is injected into the $D_2$ gas. The lack of Bremsstrahlung background, makes the Li line dominant, and so the instrumental function can be inferred directly for each channel. The interference filter curve is measured separately by putting a uniform intensity light emission diode (LED) in front of the spectrometer. Both the instrumental function and the interference filter curve must be determined for each channel since they can be different for different channels due to fiber geometry, lens contamination, etc.

For constructing the likelihood, we need to model the uncertainties in the system. There are two major sources of uncertainties: 1) Poisson noise $\sigma_{ph}=\sqrt{n_{ph}}$ from the photon statistics and 2) the electronic noise $\sigma_e$. For Poisson noise the number of measured photons, $n_{ph}$, is calculated from the number of photoelectrons in the signal using the CCD camera's photons to photoelectron ratio. When the number of photons is sufficiently large the Poisson noise can be approximated by a Gaussian distribution. The variance of the Gaussian distributed electronic noise is estimated from measurements without exposure, i.e., the fluctuation level of the background signal. This gives the following likelihood
\begin{equation}
p\lp\mathbf{y}|\mathbf{y}_*\rp=\frac1{\sqrt{\lp2\pi\rp^N\abs{\mathbf{\Sigma}}}}\exp{\lp-\frac12\lp\mathbf{y}-\mathbf{y}_*\rp^T\mathbf{\Sigma}^{-1}\lp\mathbf{y}-\mathbf{y}_*\rp\rp},
\label{eq:likelihood-system}
\end{equation}
where the $N\times N$ diagonal matrix $\mathbf{\Sigma}$ provides the associated uncertainties for $N$ pixels in a channel defined as $\mathbf{\Sigma}=\mathbf{\Sigma}_{ph}+\mathbf{\Sigma}_e$. $\mathbf{\Sigma}_{ph}$ and $\mathbf{\Sigma}_e$ are
\begin{eqnarray}
\mathbf{\Sigma}_{ph}&=&
\lsb \begin{array}{cccc}
\sigma^2_{ph1} & & & \\
& \sigma^2_{ph2} & & \\
& & \dots & \\
& & & \sigma^2_{phN} \\
\end{array}\rsb,
\nonumber\\
\mathbf{\Sigma}_{e}&=&
\lsb \begin{array}{cccc}
\sigma_{e1}^2 & & & \\
& \sigma_{e2}^2 & & \\
& & \dots & \\
& & & \sigma_{eN}^2 \\
\end{array}\rsb,
\label{eq:noise}
\end{eqnarray}
where the subscript $\lcb1,2,\dots,N\rcb$ corresponds to the pixel index of a channel in the CCD camera. Notice that the only unknown in \refeq{eq:likelihood-system} is $\vecy_*$.

The prior $p\lp\vecy_*\rp$, defined in \refeq{eq:GPR-explicit} contains the three hyperparameters $\sigma_f$, $\sigma_n$ and $l$. Since the noise of the signal is captured by the likelihood, we do not need to include the noise variance $\sigma_n$ in the prior \refeq{eq:cov}. However, for the sake of numerical stability we choose $\sigma_n$ such that $\sigma_n/\sigma_f\ll 1$, i.e., $\sigma_n^2/\sigma_f^2=10^{-2}$ for both the instrumental function and the interference filter curve. The hyperparameters, both the overall scale $\sigma_f$ and the length scale $l$, are determined by maximizing the evidence $p\lp\vecy\rp$ in the denominator of Bayes formula \refeq{eq:Bayes}. We choose the hyperparameters such that $p\lp\vecy|\sigma_f,l\rp$ is maximized, where $p\lp\vecy|\sigma_f,l\rp$ can be found from the marginalization of the joint distribution of data and free parameters
\begin{equation}
p\lp\vecy|\sigma_f,l\rp=\int p\lp\vecy|\vecy_*,\sigma_f,l\rp p\lp\vecy_*|\sigma_f,l\rp d\vecy_*.
\label{eq:evidence}
\end{equation}

\begin{figure}
\includegraphics[width=\linewidth]{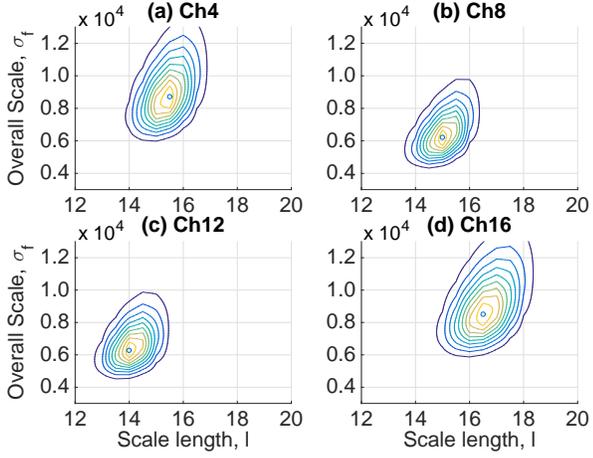}
\caption{Contour of the evidence probability density for the interference filter curve calculated by \refeq{eq:evidence} as a function of the length scale $l$ and the overall scale $\sigma_f$ for the spatial channel of (a) $m=4$, (b) $8$, (c) $12$ and (d) $16$. Both the overall scale and the length scale of the corresponding spatial channel of the interference filter curve in \refeq{eq:cov} are the values that maximize the evidence.}
\label{fig:evidence-filter}
\end{figure}

\begin{figure}
\includegraphics[width=\linewidth]{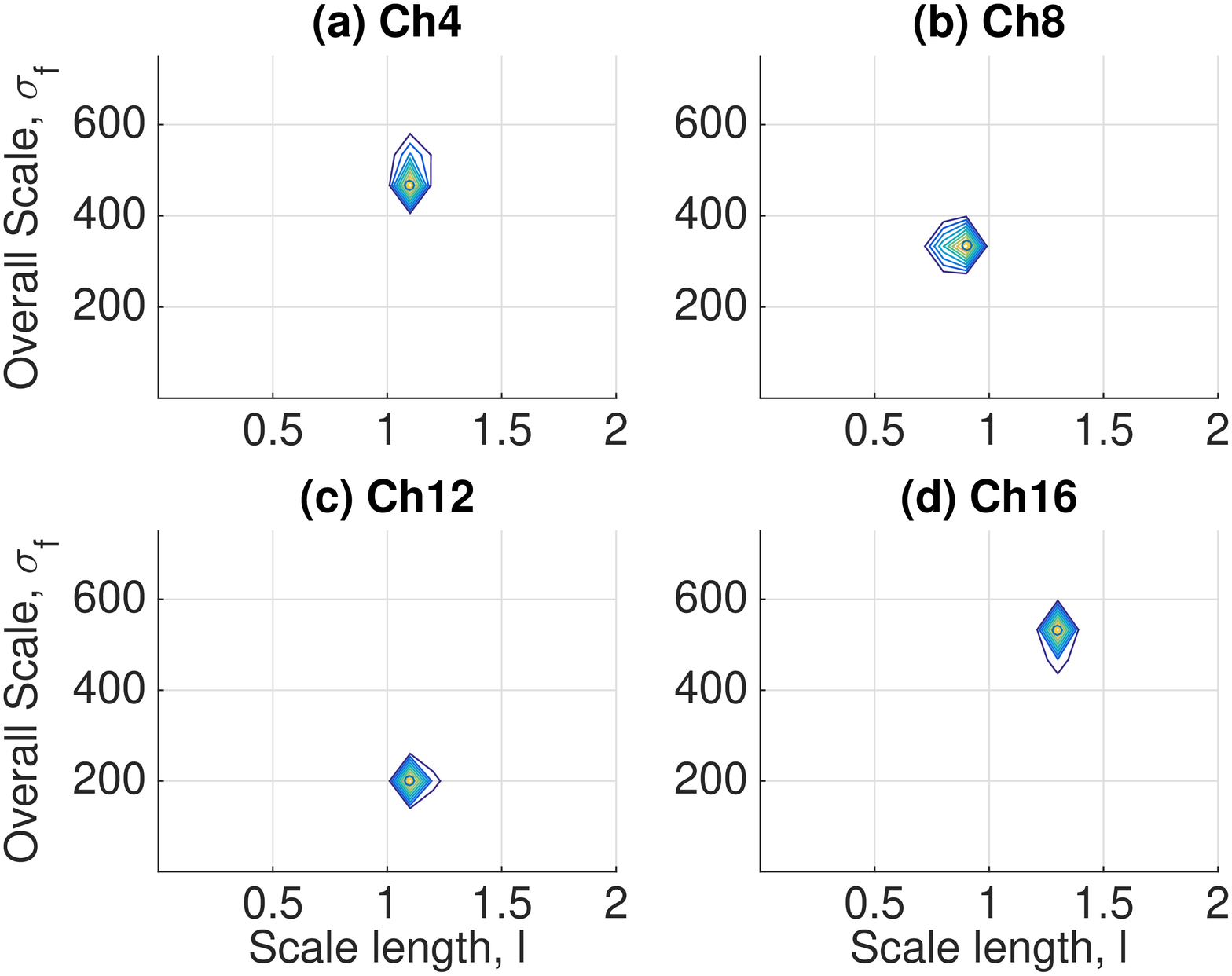}
\caption{Same as \reffig{fig:evidence-filter} for the instrumental function.}
\label{fig:evidence-line}
\end{figure}

\refapp{app:evidence} discusses the rationale behind choosing hyperparameters that maximizes the evidence. Note that the right hand side of \refeq{eq:evidence} contains the likelihood and the prior for which we have well defined expressions, \refeq{eq:likelihood-system} and \refeq{eq:GPR-explicit}, respectively. \reffig{fig:evidence-filter} and \reffig{fig:evidence-line} show the $p\lp\vecy|\sigma_f,l\rp$ of interference filter curves and instrumental function, respectively, as a function of the scale length $l$ and the overall scale $\sigma_f$ for the spatial channel of (a) $m=4$, (b) $8$, (c) $12$ and (d) $16$.

\begin{figure}
\includegraphics[width=\linewidth]{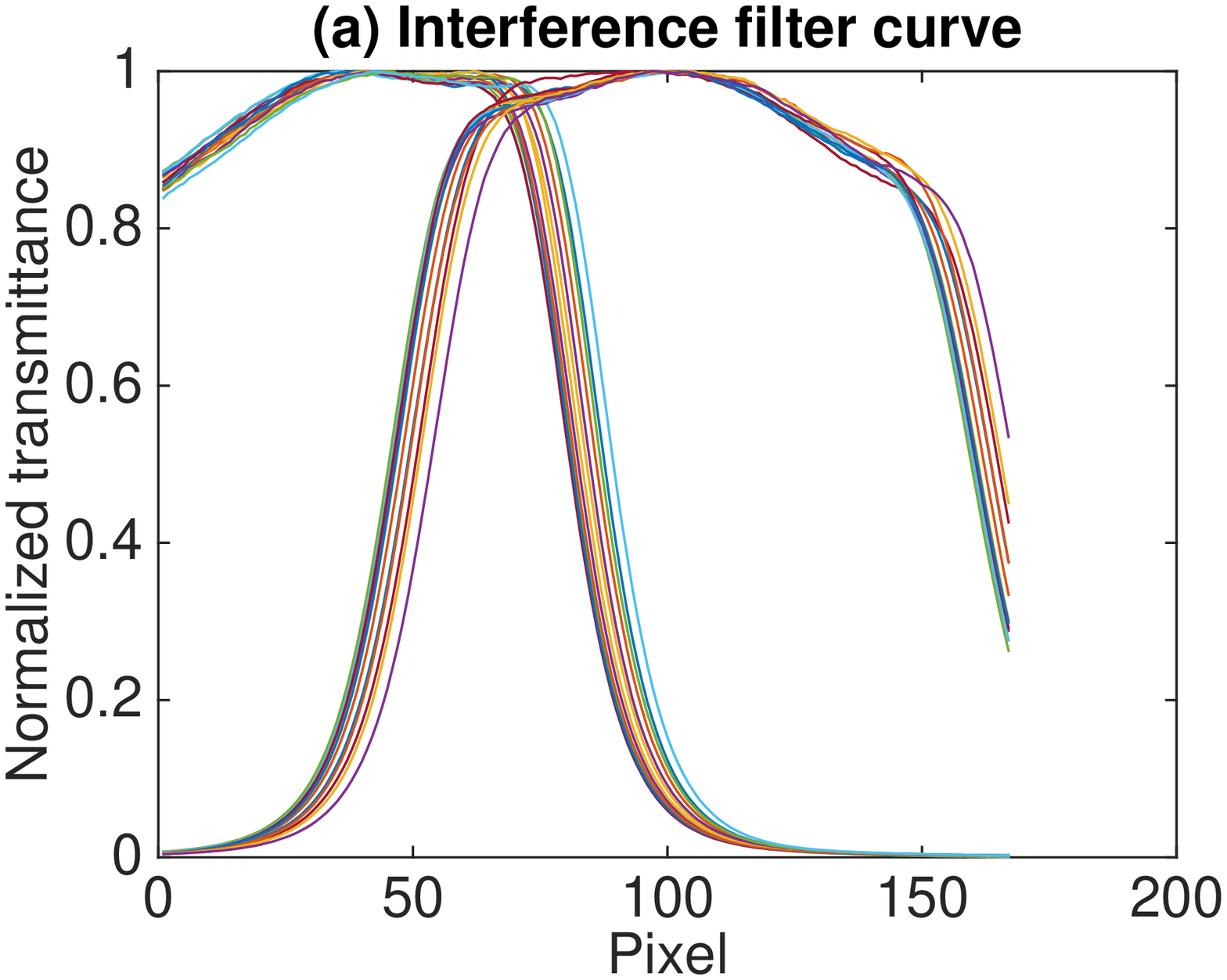}
\includegraphics[width=\linewidth]{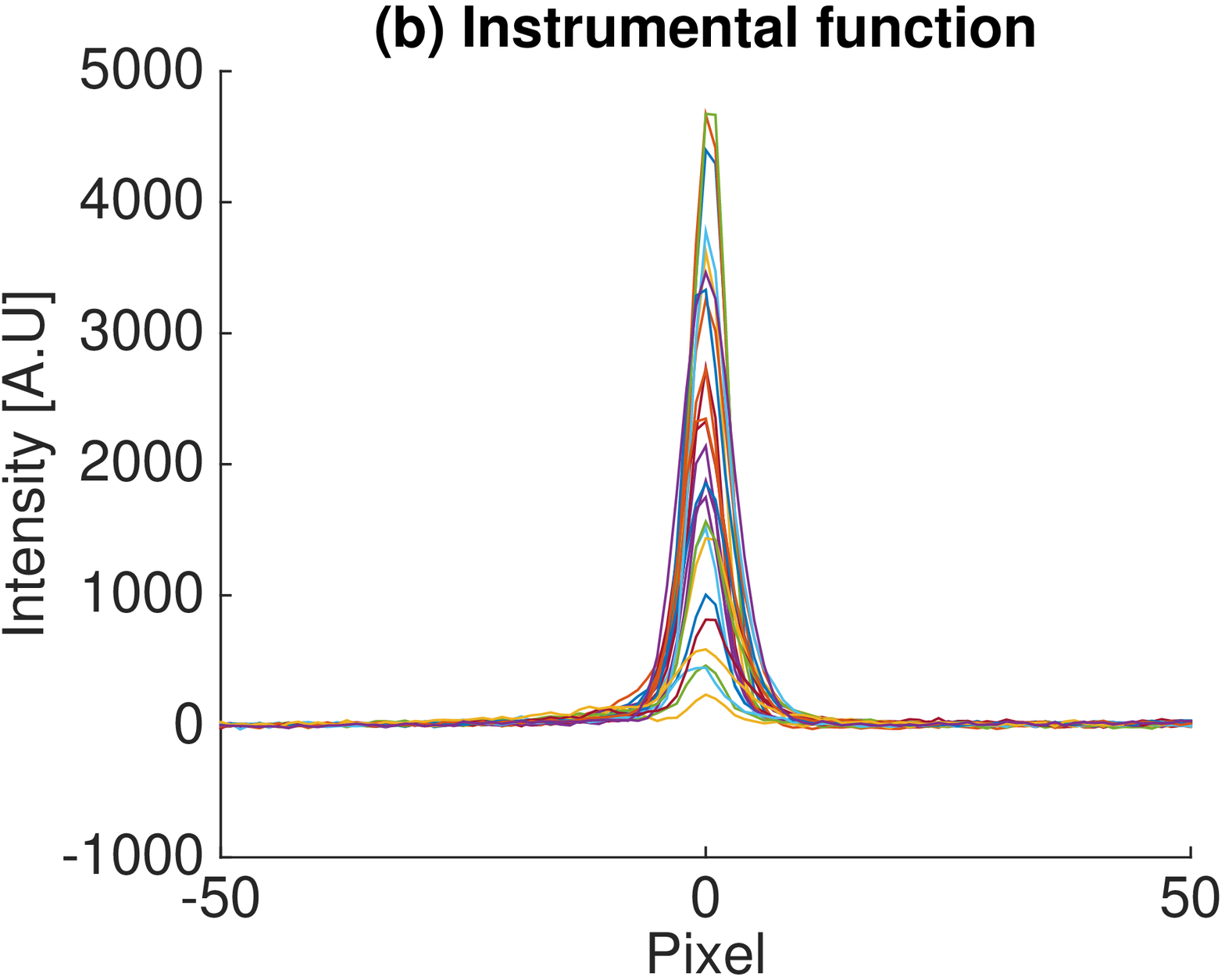}
\caption{Best estimate (a) normalized interference filter curves and (b) instrumental functions as a function of pixel of the CCD camera. Different colors correspond to different channels of the JET Li-BES system. The pixel can be converted to wavelength, the dispersion being approximately $0.04$ nm/pixel.}
\label{fig:resultss-system}
\end{figure}

Since both the prior and the likelihood are multivariate Gaussian, and the forward model is linear, the posterior distribution is also a multivariate Gaussian distribution over $\vecy_*$. The posterior mean and covariance can thus be calculated explicitly via a Bayesian linear Gaussian inversion.\cite{08PPCF_Svensson} \reffig{fig:resultss-system} shows (a) the normalized interference filter curve $d_m\lp x\rp$ and (b) the instrumental function $c_m\lp x\rp$ as a function of pixel of the CCD camera. Each pixel ($x$) corresponds to a specific wavelength where the wavelength calibration is performed with neon and xenon lamps.\cite{10RSI_Brix}

\subsection{Intensity profile inference}
\label{sec:inference-line}
In this part, we obtain the intensities in \refeq{eq:spectrum}, i.e., the intensity of Li line $a_m$, Bremsstrahlung background $b_m$ and instrumental offset $z_m$ from the JET Li-BES data of each channel. Again, Bayesian probability theory is utilized to determine these three quantities:
\begin{equation}
p\lp\calI_m|\calD_m\rp\propto p\lp\calD_m|\calI_m\rp p\lp\calI_m\rp,
\label{eq:Bayes-line}
\end{equation}
where $\calI_m=\lsb a_m, b_m, z_m \rsb$ is a vector of the free parameters for the $m^{\mathrm{th}}$ channel, and $\calD_m$ is the measured data. 

As priors for $a_m$, $b_m$ and $z_m$ we use
\begin{eqnarray}
p\lp a_m\rp&=&\frac1{\sigma_{a_m}\sqrt{2\pi}}\exp\lsb-\frac{\lp a_m-0\rp^2}{2\sigma_{a_m}^2}\rsb
\nonumber\\
p\lp b_m\rp&=&\frac1{\sigma_{b_m}\sqrt{2\pi}}\exp\lsb-\frac{\lp b_m-0\rp^2}{2\sigma_{b_m}^2}\rsb
\nonumber\\
p\lp z_m\rp&=&\frac1{\sigma_{z_m}\sqrt{2\pi}}\exp\lsb-\frac{\lp z_m-0\rp^2}{2\sigma_{z_m}^2}\rsb,
\label{eq:prior}
\end{eqnarray}
giving $p\lp\calI_m\rp=p\lp a_m\rp p\lp b_m\rp p\lp z_m\rp$. We choose very large prior standard deviations $10^6$, making the Gaussians effective flat. The likelihood $p\lp\calD_m|\calI_m\rp$ is multivariate Gaussian
\begin{eqnarray}
p\lp\calD_m|\calI_m \rp&=&\frac{1}{\sqrt{\lp2\pi\rp^N}\abs{\mathbf{\Sigma}}}\times\\
&&\exp\lp-\frac{1}{2}\lp \calD_m-s_m\rp^T\mathbf{\Sigma}^{-1}\lp \calD_m-s_m\rp \rp\nonumber, 
\label{eq:likelihood-spectrum}
\end{eqnarray}
where the covariance matrix $\mathbf{\Sigma}$ is given by \refeq{eq:noise} and $s_m$ by \refeq{eq:spectrum} which is a function of $\calI_m$.

\begin{figure}
\includegraphics[width=\linewidth]{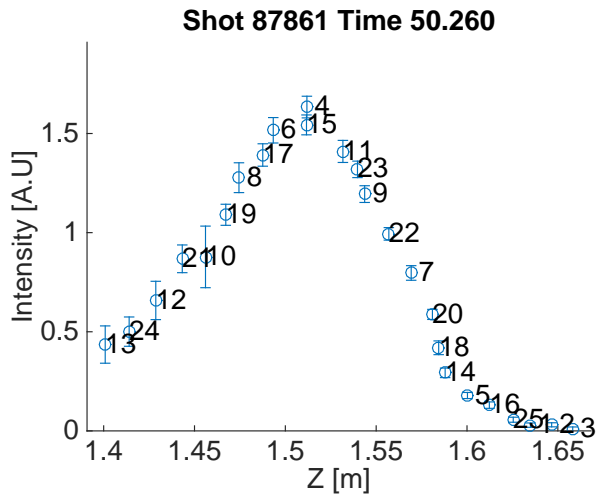}
\caption{The intensity of lithium line radiation $a_m$, profile as a function of $Z$ (distance from the midplane) with channel numbers. Circles are the intensities at their maximum posterior, and vertical bars represent $3\sigma$ ranges.}
\label{fig:results}
\end{figure}

Having a well defined prior and likelihood, the posterior distribution is obtained by a Bayesian linear Gaussian inversion. \reffig{fig:results} shows the intensity of lithium line radiation $a_m$, profile as a function of $Z$, the distance of the channel location from the midplane (\reffig{fig:system}). Here we only have $25$ channels since the interference filter curve of one channel could not be measured due to a technical problem. The numbers in the figure are the channel numbers of the JET Li-BES system. Note that this intensity profile is subsequently used to reconstruct the electron density profile at the plasma edge. We do not include inference of the electron density profile since it is outside the scope of this paper.

\begin{figure}
\includegraphics[width=\linewidth]{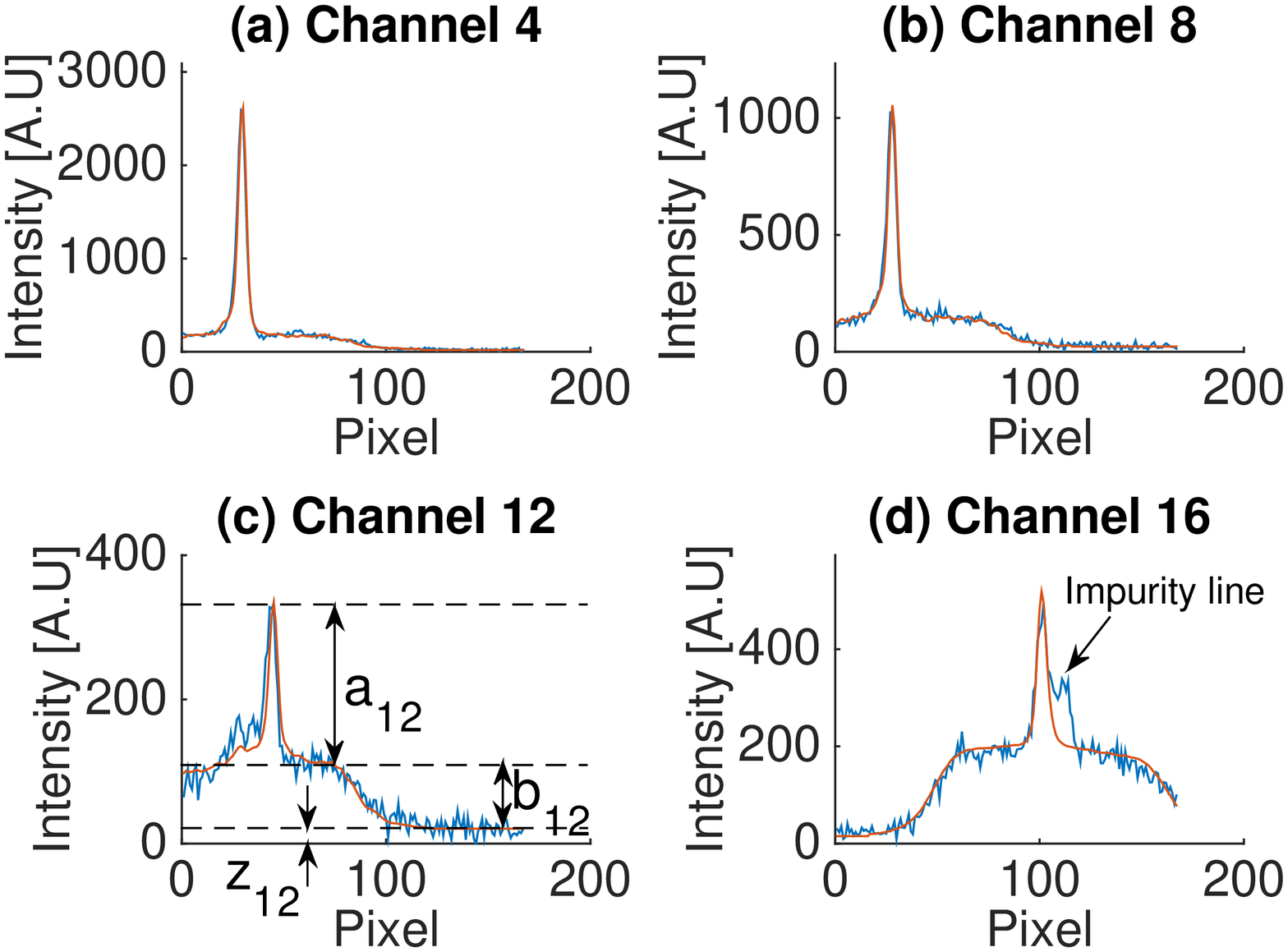}
\includegraphics[width=\linewidth]{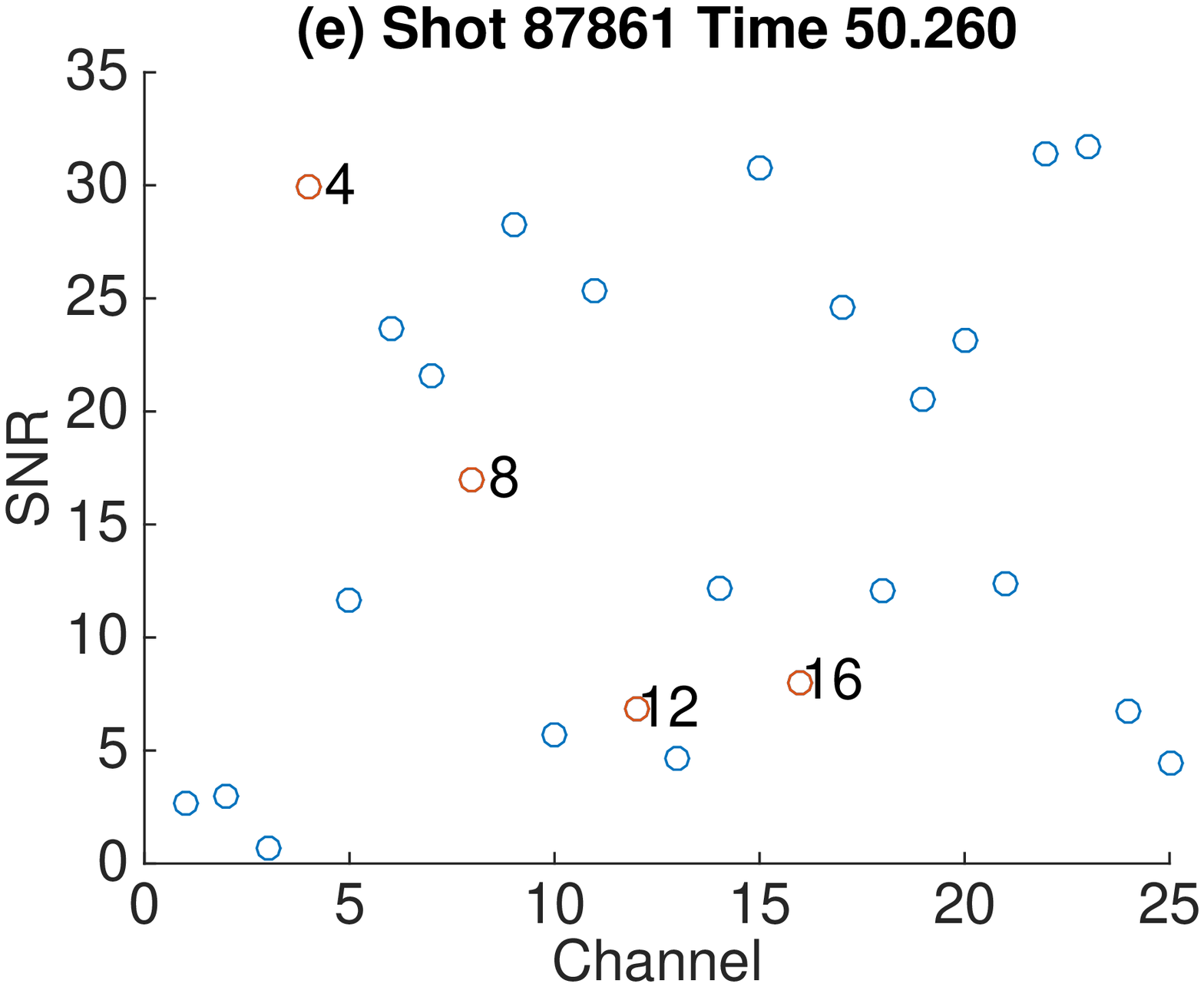}
\caption{Estimated spectrum $s_m$ (red) as a function of pixel (wavelength) for channel numbers  (a) $m=4$, (b) $8$, (c) $12$ and (d) 16 and the measured spectra (blue). An example of estimated intensity of lithium line radiation $a_m$, background radiation $b_m$ and offset $z_m$ for $m=12$ is shown in (c). Signal-to-noise ratio of each channel is shown in (e). The red circles in (e) are the channels shown in (a)-(d).}
\label{fig:results-spectrum}
\end{figure}

Having determined all the quantities on the right hand side of \refeq{eq:spectrum} in a consistent way based on Bayesian inference and Gaussian processes, in  \reffig{fig:results-spectrum}(a)-(d) shows the estimated spectra $s_m\lp x\rp$ (red) and the measured data (blue) for channel numbers $m=4$, $8$, $12$ and $16$. They show that the lithium line shape and its absolute intensity, background radiation and electronic offsets are well reconstructed. In \reffig{fig:results-spectrum}(c), we illustrate the intensity of lithium line radiation $a_{12}$, background radiation $b_{12}$ and offset $z_{12}$ for $m=12$. For the case of channel $m=16$, i.e., \reffig{fig:results-spectrum}(d), there exists an extra peak in the measurement caused by impurity line radiation.\cite{10RSI_Brix} The estimated $s_m$ does not capture the impurity lines as they are not included in our model. As a simple consistency check on estimated uncertainties of $a_m$ in \reffig{fig:results}, we can see that the larger the signal-to-noise ratio, the higher the intensity of lithium line radiation $a_m$ as shown in \reffig{fig:results-spectrum}(e).

\section{Summary}
\label{sec:summary}
As Li-BES systems are widely used to reconstruct the electron density profile at the edge of plasmas based on the intensity of Li line, being able to resolve the measured spectrum data into the intensity of Li line, Bremsstrahlung background and instrumental offset is a substantial improvement on the conventional method of using background subtraction through beam modulation, and a Gaussian fit of the line shape. To obtain these parameters, we also need to know the instrumental function and interference filter curve.

The instrumental function and interference filter curve are both modelled with Gaussian processes, separately for each channel. The length scales and overall scales of the curves have been determined directly from the data through maximization of the evidence. The reconstructed spectra agree well with the measured spectra. In addition, the associated uncertainties of the data are also obtained consistently.

Apart from the improvement in fitting quality, another major advantage of this approach is that independent measurements of the background signals do not need to be done. The method of background measurements is done by modulating the beam with an electrostatic deflection plate which increases hardware complexity and loses some temporal information. The Bremsstrahlung background signal that is simultaneously inferred could give additional information on for example effective charge.

\begin{acknowledgments}
This work is supported by National R\&D Program through the National Research Foundation of Korea (NRF) funded by the Ministry of Science, ICT \& Future Planning (grant number 2014M1A7A1A01029835) and the KUSTAR-KAIST Institute, KAIST, Korea. 
This work has been carried out within the framework of the EUROfusion Consortium and has received funding from the Euratom research and training programme 2014-2018 under grant agreement No 633053. The views and opinions expressed herein do not necessarily reflect those of the European Commission.
\end{acknowledgments}

\appendix
\section{Rationale behind maximizing the evidence probability}
\label{app:evidence}
The posterior for the instrumental function or interference filter curve can be written as 
\begin{eqnarray}
p\lp\vecy_*|\vecy\rp&\propto&\exp{\lsb-\frac12(\mathbf{y}-\mathbf{y}_*)^T\mathbf{\Sigma}^{-1}(\mathbf{y}-\mathbf{y}_*)\rsb}\times
\nonumber\\
&&\exp\lsb-\frac{1}{2}\lp\vecy_*-\vecO\rp^T\matK^{-1}\lp\vecy_*-\vecO\rp\rsb,
\label{eq:posterior}
\end{eqnarray}
using \refeq{eq:likelihood-system} and \refeq{eq:GPR-explicit}. In addition to our main unknown $\vecy_*$, note that $\matK$ is a function of the hyperparameters $\sigma_f$ and $l$, which are also not known. This suggests that we should integrate out these hyperparameters to get the marginal posterior:
\begin{eqnarray}
p\lp\vecy_*|\vecy\rp&=&\int\int p\lp\vecy_*,\sigma_f,l|\vecy\rp d\sigma_fd\l
\nonumber\\
&=&\int\int p\lp\vecy_*|\sigma_f,l,\vecy\rp p\lp\sigma_f,l|\vecy\rp d\sigma_fdl
\nonumber\\
&\approx&p\lp\vecy_*|\sigma_{f0},l_0,\vecy\rp,
\label{eq:posterior-explicit}
\end{eqnarray}
where the approximation is realized by setting $p\lp\sigma_f,l|\vecy\rp=\delta\lp\sigma_f-\sigma_{f0}\rp\delta\lp l-l_0\rp$, which is valid if the posterior distribution over the hyperparameters is narrowly centered around their most probable values.

The posterior for the hyperparameters $\sigma_{f0}$ and $l_0$ is given by
\begin{equation}
p\lp \sigma_f,l|\vecy\rp\propto p\lp\vecy|\sigma_f,l\rp p\lp\sigma_f,l\rp,
\end{equation}
where $p\lp\sigma_f,l|\vecy\rp$ is given by \refeq{eq:evidence}. Using a uniform prior over $\sigma_f$ and $l$, the maximum posterior corresponds to the maximum of the marginal likelihood (\refeq{eq:evidence}).

\bibliography{RSI2015}

\end{document}